\documentclass[aps,prl,reprint,groupedaddress,showpacs]{revtex4-1}
\usepackage{graphicx}
\usepackage{dcolumn}
\usepackage{bm}
\usepackage[latin1]{inputenc}
\usepackage{longtable}

\begin{document}

\title{Fresnel coefficients of a two-dimensional atomic crystal}

\author{Michele Merano}
\email[]{michele.merano@unipd.it}

\affiliation{Dipartimento di Fisica e Astronomia G. Galilei, Universit$\grave{a}$ degli studi di Padova, via Marzolo 8, 35131 Padova, Italy}

\date{\today}

\begin{abstract}
In general the experiments on the linear optical properties of a single-layer two-dimensional atomic crystal are interpreted by modeling it as a homogeneous slab with an effective thickness. Here I fit the most remarkable experiments in graphene optics by using the Fresnel coefficients, fixing both the surface susceptibility and the surface conductivity of graphene. It is shown that the Fresnel coefficients and the slab model are not equivalent. Experiments indicate that the Fresnel coefficients are able to simulate the overall experiments here analyzed, while the slab model fails to predict absorption and the phase of the reflected light.
\end{abstract}

\maketitle
\section{}
Graphene, a two-dimensional (2D) carbon-based crystal, was the first atomically thin layer of atoms that was produced \cite{Novoselov2004}. The family of 2D crystals however is not limited to graphene. Boron nitride or the transition-metal dichalcogenides have been grown as atomic monolayers \cite{Novoselov2005}. The characteristics of the 2D materials might be very different from those of their 3D precursors. This is true also for their optical properties. For example, it was argued that the high-frequency conductivity ($\sigma$) in graphene should be a universal constant equal to $e^{2}/4\hbar$ \cite{Ando2002}. This is due to the exactly zero overlapping between the valence and the conduction bands in graphene (a zero-gap semiconductor), while it is finite in graphite. Atomically thin transition-metal dichalcogenides are direct band semiconductors \cite{Heinz2010} while the bulk materials have got an indirect band gap.

Also the optical reflectivity, transmissivity, and absorption of a 2D crystal are remarkable. The absorption of graphene is determined by the fine-structure constant \cite{Nair2008}. Microfabrication of devices used in many experimental studies currently relies on the fact that 2D crystals can be visualized using optical microscopy if prepared on top of suitable substrates \cite{Blake2007, Kis11, Sandhu13, Rhee12, Hansen11, Rabe10, Dai15, Nolte09, Rubio10}. This is due to a multiple-reflection effect that enhances the visibility of the atomically thin layer, where the optical properties of the 2D crystal play a fundamental role. Measurements of the optical dielectric functions of 2D crystals by spectroscopic ellipsometry have already been reported \cite{Kravets2010, Heinz2014, Diebold10, Weiss10, Gajic15, Sanden10}. The interpretation of these results relies on a model of the monolayer treated as a homogeneous medium with an effective thickness \cite{Blake2007, Kis11, Sandhu13, Rhee12, Hansen11, Rabe10, Dai15, Nolte09, Rubio10, Kravets2010, Heinz2014, Diebold10, Weiss10, Gajic15, Sanden10}. 

An alternative to this model is to treat the 2D crystal as a boundary and to provide the right boundary conditions for a single atomic layer. This approach has been used in different papers to derive the reflection and the transmission coefficients of light between two media separated by a graphene flake \cite{Pershoguba07, Hanson08, Stauber2008, Zhan2013, Galina15}. No reflection or transmission coefficients for other 2D materials different from graphene have been considered until now. Although the approach used in \cite{Pershoguba07, Hanson08, Stauber2008, Zhan2013, Galina15} is substantially correct, these papers are primarily interested in the microscopic theoretical description of graphene. In general a comparison with experimental results of the reflection and transmission coefficients so obtained was limited to absorption and it was not extended to other experiments such as optical contrast or ellipsometry. 

Here I deduce an expression for the Fresnel coefficients valid for any single-layer 2D atomic crystal. I adopt a classical macroscopic approach. I consider first the case of a freestanding non-conducting material (such as boron nitride \cite{Blake2011}) and then I will turn my attention to optics of conducting media (such as graphene). Then will be the turn of 2D crystals on substrates. The formulas obtained are functional to a comparison with published experimental results on graphene, by far the most studied 2D crystal, and they will enlighten the power of Fresnel coefficients for simulating the overall phenomena described above.

Consider a flat 2D crystal, composed of $N$ atoms per square cm with an atomic polarizability $\alpha$ \cite{Purcell}. If we apply an electric field in the plane of the crystal a macroscopic dipole moment arises and it is possible to define a density of polarization $\vec{\textbf{\emph{P}}}$. If the electric field is applied orthogonally to the 2D crystal, no macroscopic polarization is created (or in any case it is much smaller and I do not consider it). Indeed to have a macroscopic polarization the microscopic dipoles need to be aligned, to generate a macroscopic separation of charges. As a further simplification I suppose that the 2D crystal is isotropic in its own plane. This seems realistic because graphite, for instance, is a uniaxial crystal with the optical axis along the graphene's exfoliation axis. I assume also that $\vec{\textbf{\emph{P}}}=\epsilon_{0}\chi\vec{\textbf{\emph{E}}}$ where $\epsilon_{0}$ is the vacuum permittivity and $\chi$ is the electric susceptibility. Wherever the polarization in matter changes with time there is an electric current $\vec{\textbf{J}}_{p}$, a genuine motion of charges. The connection between rate of change of polarization and current density is $\vec{\textbf{J}}_{p}=\partial{\vec{\emph{\textbf{P}}}}/\partial{t}$. It is important to note that in passing from a bulk to a 2D crystal the dimensions of $\vec{\textbf{\emph{P}}}$ pass from dipole moment/volume to dipole moment/area, the dimensions of $\chi$ pass from a pure number to a length and $\vec{\textbf{J}}_{p}$ is a current per unit length. It is not the aim of this paper to give a microscopic theory that furnishes $\chi$. Experiments will fix it.
\begin{figure}
\includegraphics{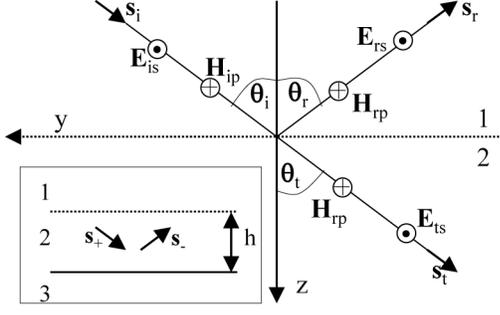}
\caption{\label{} Oblique incidence of a plane wave on graphene. The electric (magnetic) field for $s$ ($p$) polarization is shown. Inset: Three layer substrate; graphene is deposited at the interface of media 1 and 2.}
\end{figure}

Suppose that a 2D crystal is suspended in vacuum and that a plane wave falls onto it (Fig. 1; $n_{1}=n_{2}=1$); the relation in between $\vec{\emph{\textbf{E}}}$ and $\vec{\emph{\textbf{H}}}$ in the incident, reflected and transmitted waves is $\eta\vec{\textbf{\emph{H}}}=\hat{s}\wedge\vec{\textbf{\emph{E}}}$ where $\hat{s}$ is the unit vector in along the propagation direction and $\eta$ is the impedance of vacuum. The boundary conditions are $\hat{\kappa} \wedge (\vec{\textbf{E}}_{2}-\vec{\textbf{E}}_{1})=0$, $\hat{\kappa} \wedge (\vec{\textbf{H}}_{2}-\vec{\textbf{H}}_{1})=\textbf{J}_{p}$ where $\hat{\kappa}$ is the unit vector along the $z$ axis. For $s$ ($p$) polarization then, 
\begin{eqnarray}
E_{xi}+E_{xr}&=&E_{xt};\qquad \qquad (E_{yi}-E_{yr}=E_{yt}) \\
E_{xi}+E_{xr}&=&\frac{P_{x}}{\epsilon_{0}\chi};\qquad \qquad (E_{yi}-E_{yr}=\frac{P_{y}}{\epsilon_{0}\chi}) \nonumber  \\
H_{yi}-H_{yr}&=&H_{yt}+i\omega P_{x};\quad (H_{xi}+H_{xr}=H_{xt}+i\omega P_{y})  \nonumber
\end{eqnarray}
where time dependence $\exp(i\omega t)$ ($\omega$ is the angular frequency of the light) has been assumed.

Defining \cite{Landau} $r_{s}=E_{r}/E_{i}$, $t_{s}=E_{t}/E_{i}$ and $r_{p}=H_{r}/H_{i}$, $t_{p}=H_{t}/H_{i}$ as the reflection and the transmission coefficients, we have
\begin{eqnarray}
r_{s}=\frac{-ik\chi}{ik\chi+2\cos\theta_{i}}; \qquad r_{p}=\frac{ik\chi\cos\theta_{i}}{ik\chi\cos\theta_{i}+2}
\end{eqnarray}
and $t_{s}=r_{s}+1$, $t_{p}=1-r_{p}$ ($k$ is the wave vector of light in vacuum and $\theta_{i}$ is the angle of incidence). From energy flux considerations \cite{Wolf}, in this special case, the reflectivity  is $R_{s(p)}=|{r}^{2}_{s(p)}|$, the transmissivity is $T_{s(p)}=|{t}^{2}_{s(p)}|$, and their sum $R_{s(p)}+T_{s(p)}=1$ shows that there is no absorption.

We turn now our attention to conducting media. The Ohm's law $\vec{\textbf{J}}=\sigma\vec{\emph{\textbf{E}}}$ must be taken into account; again we assume that $\vec{\textbf{J}}$ can not exist (or it is small) in a direction orthogonal to the crystal plane, and in-plane isotropy. The boundary conditions for $\vec{\emph{\textbf{H}}}$ changes into $\hat{\kappa} \wedge (\vec{\textbf{H}}_{2}-\vec{\textbf{H}}_{1})=\textbf{J}_{p}+\textbf{J}$, and we add to Eqs. (1) the Ohm's law:
\begin{eqnarray}
E_{xi}+E_{xr}= \frac{j_{x}}{\sigma}; \quad (E_{yi}-E_{yr}= \frac{j_{y}}{\sigma})
\end{eqnarray}
for $s$ ($p$) polarization.
We obtain
\begin{eqnarray}
r_{s}=-\frac{ik\chi+\sigma\eta}{ik\chi+\sigma\eta +2\cos\theta_{i}}; r_{p}=\frac{(ik\chi+\sigma\eta)\cos\theta_{i}}{(ik\chi+\sigma\eta)\cos\theta_{i}+2}\qquad
\end{eqnarray}
and $t_{s}=r_{s}+1$, $t_{p}=1-r_{p}$. As for a bulk material conductivity is connected with the transformation of part of the electromagnetic energy into heat. At normal incidence,
\begin{eqnarray}
T_{s(p)}+R_{s(p)}=1-\frac{4\sigma\eta}{4+4\sigma\eta+\sigma^{2}\eta^{2} + k^{2}\chi^{2}} \simeq1-\sigma\eta  \nonumber
\end{eqnarray}
In the case of graphene, at wavelengths $\lambda$ where $\sigma=e^{2}/4\hbar$, we retrieve the remarkable result that the fine-structure constant defines its optical transparency \cite{Nair2008}.

The reflection and the transmission coefficient for the case of a 2D crystal at the interface of two different bulk media (1 and 2 in Fig. 1)  is now easily obtained. Only the relation in between $\vec{\emph{\textbf{E}}}$ and $\vec{\emph{\textbf{H}}}$ in the incident, reflected, and transmitted waves changes:
\begin{eqnarray}
\frac{\eta}{n_{1}}\vec{\textbf{\emph{H}}}_{i(r)}=\hat{s}_{i(r)}\wedge\vec{\textbf{\emph{E}}}_{i(r)}; \quad \frac{\eta}{n_{2}}\vec{\textbf{\emph{H}}}_{t}=\hat{s}_{t}\wedge\vec{\textbf{\emph{E}}}_{t};
\end{eqnarray}
where $n_{1}$, $n_{2}$ are the refractive indexes. We obtain:
\begin{eqnarray}
r_{s}&=&\frac{n_{1}\cos\theta_{i}-n_{2}\cos\theta_{t}-ik\chi-\sigma\eta}{n_{1}\cos\theta_{i}+n_{2}\cos\theta_{t}+ik\chi+\sigma\eta }\\ r_{p}&=&\frac{n_{2}\cos\theta_{i}-n_{1}\cos\theta_{t}+(ik\chi+\sigma\eta)\cos\theta_{i}\cos\theta_{t}}{n_{2}\cos\theta_{i}+n_{1}\cos\theta_{t}+(ik\chi+\sigma\eta)\cos\theta_{i}\cos\theta_{t}} \nonumber
\end{eqnarray}
and $t_{s}=r_{s}+1$, $t_{p}=(1-r_{p})n_{2}\cos\theta_{i}/(n_{1}\cos\theta_{t})$.

The reflection coefficient of a 2D crystal on a stratified medium is of primary importance also. I consider the case represented in Fig. 1 (inset), corresponding to published experimental work that will be considered in this article. The boundary conditions for the interface in between bulk media 2 and 3 are $\hat{\kappa} \wedge (\vec{\textbf{E}}_{3}-\vec{\textbf{E}}_{2})=0$, $\hat{\kappa} \wedge (\vec{\textbf{H}}_{3}-\vec{\textbf{H}}_{2})=0$. Equation (5) is easily extended to medium 3 \cite{Wolf}. For $s$ polarization,
\begin{eqnarray}
&E&_{xi}+E_{xr}=E_{x+}+E_{x-}; \qquad E_{xi}+E_{xr}=\frac{P_{x}}{\epsilon_{0}\chi}   \\
&E&_{xi}+E_{xr}= \frac{j_{x}}{\sigma}; \quad H_{yi}-H_{yr}=H_{y+}-H_{y-}+i\omega P_{x}+J_{x} \nonumber \\
&E&_{x+}e^{-i\beta}+E_{x-}e^{i\beta}=E_{xt}; \quad H_{y+}e^{-i\beta}-H_{y-}e^{i\beta}=H_{yt}; \nonumber
\end{eqnarray}
and for $p$ polarization,
\begin{eqnarray}
&E&_{yi}-E_{yr}=E_{y+}-E_{y-}; \qquad E_{yi}-E_{yr}=\frac{P_{y}}{\epsilon_{0}\chi}   \\
&E&_{yi}-E_{yr}= \frac{j_{y}}{\sigma}; \quad H_{xi}+H_{xr}=H_{x+}+H_{x-}+i\omega P_{y}+J_{y} \nonumber \\
&E&_{y+}e^{-i\beta}-E_{y-}e^{i\beta}=E_{yt}; \quad H_{x+}e^{-i\beta}+H_{x-}e^{i\beta}=H_{xt}; \nonumber
\end{eqnarray}
where $\beta=kn_{2}hcos\theta_{2}$, $\theta_{2}$ is the propagation angle in medium 2, and $h$ is its thickness \cite{Wolf}. This furnishes
\begin{eqnarray}
r_{s}&=&\frac{r_{12s}+r_{23s}e^{-2i\beta}-t_{12s}\frac{ik\chi+\sigma\eta}{2\cos\theta_{i}}(1+r_{23s}e^{-2i\beta})}{1+r_{12s}r_{23s}e^{-2i\beta}+t_{12s}\frac{ik\chi+\sigma\eta}{2\cos\theta_{i}}(1+r_{23s}e^{-2i\beta})}\\
r_{p}&=&\frac{r_{12p}+r_{23p}e^{-2i\beta}+t_{12p}\frac{(ik\chi+\sigma\eta)\cos\theta_{2}}{2n_{2}}(1-r_{23p}e^{-2i\beta})}{1+r_{12p}r_{23p}e^{-2i\beta}+t_{12p}\frac{(ik\chi+\sigma\eta)\cos\theta_{2}}{2n_{2}}(1-r_{23p}e^{-2i\beta})}  \nonumber
\end{eqnarray}
If $\chi$ and $\sigma$ are zero we obtain the usual reflection coefficients for a three-layered medium \cite{Wolf}. The sign difference with respect to \cite{Wolf} in the face of $2i\beta$ comes from the different choice of the temporal dependence ($e^{i\omega t}$).
\begin{figure}
\includegraphics{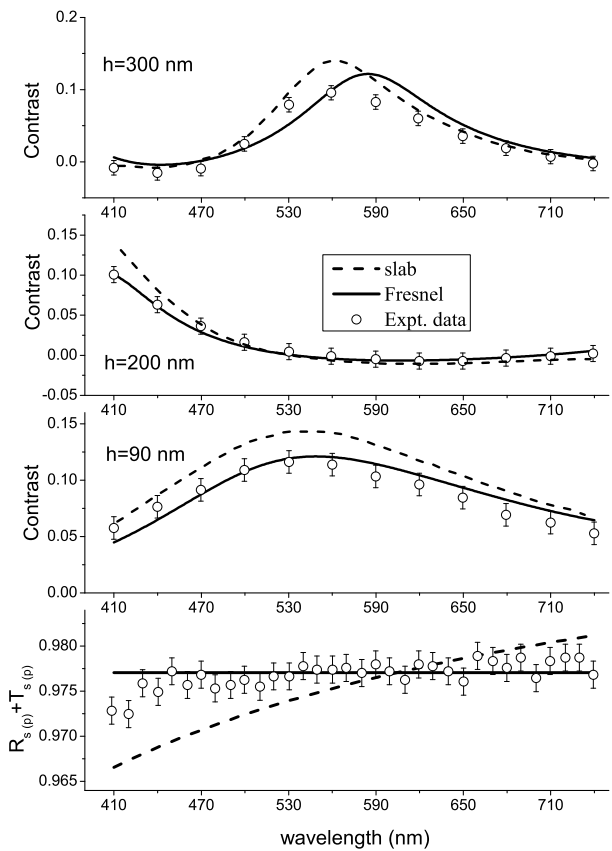}
\caption{\label{} The first three graphs show the optical contrast of graphene on Si$\rm{O_{2}}$/Si. The experimental data and the simulations based on the slab model are extracted from \cite{Blake2007}. The last graph shows the absorption of a free-standing graphene layer. Experimental data are extracted from \cite{Nair2008}.}
\end{figure}

We can now compare some of the most remarkable experiments in graphene optics with the theoretical predictions given by the Fresnel formulas (4), (6), (9) and with the predictions furnished by the slab-based model. All the experimental data reproduced in this paper have been extracted from the original papers via software digitization, ensuring a very precise reproducibility. Throughout this paper data are represented by dots, slab model predictions by a dashed line, and Fresnel model predictions by a solid line. The error bars reported in the graphs have a length of 2 standard deviations $s$ (see Table I) for the respective sets of experimental data.
 
As explained before the visibility of graphene is enhanced if prepared on top of Si$\rm{O_{2}}$/Si wafers. The contrast is defined as the relative intensity of reflected light in the presence and absence of graphene \cite{Blake2007}. Figure 2 compares the experimental data and the slab model predictions published in \cite{Blake2007} with the Fresnel theory. The first three graphs give the optical contrast for single-layer graphene on top of Si$\rm{O_{2}}$/Si wafers with three different Si$\rm{O_{2}}$ thicknesses at $\theta_{i}=0$. For the optical constants of Si$\rm{O_{2}}$ I used the Sellmeier equations, for Si I made a fit from data in \cite{Palik}. Polarizations $s$ and $p$ give the same results. The Fresnel fit reported in the first three graphs of Fig. 2 is for $\sigma=e^{2}/4\hbar=6.08\cdot 10^{-5}$ $\Omega$ and $\chi = 5\cdot 10^{-10}$ m. This is compared with the slab fit used in \cite{Blake2007}. Anyway the contrast measurements are not able to discriminate very well the value of $\chi$. All the Fresnel fits with $\sigma=e^{2}/4\hbar$ and $\chi\leq 5\cdot 10^{-10}$ m give in practice the same result. We will see that ellipsometric measurements solve this problem. 
\begin{figure}
\includegraphics{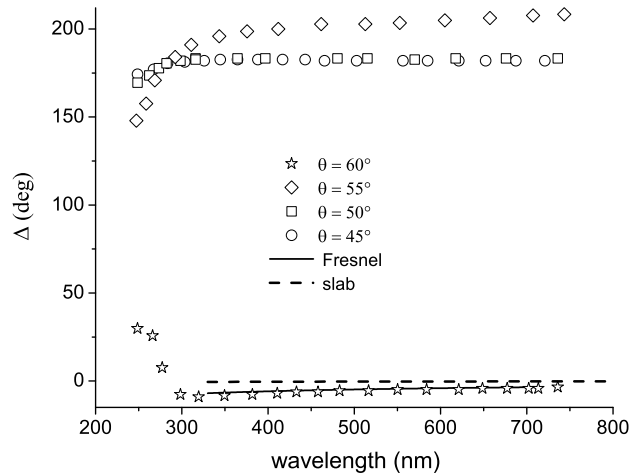}
\caption{\label{} Spectral dependance of the ellipsometric parameter $\Delta$ for graphene on an amorphous quartz substrate. Experimental data for the four $\theta_{i}$ are extracted from \cite{Kravets2010}. The Fresnel and the slab model simulations for $\theta_{i}$ =60$^\circ$ are reported (see text).}
\end{figure}

To quantify how much better the Fresnel-based model is, compared to the slab-based one, a chi-squared test is reported in Table I where $\tilde{\chi}_{0F(S)}^2$ is the reduced chi-squared value for the Fresnel (slab) fit and $P_{F(S)}$ is the probability of getting a value as large as $\tilde{\chi}_{0F(S)}^2$ assuming that the Fresnel (slab)-based model is correct \cite{Taylor}. For the contrast measurements the standard deviation $s$ has been chosen as the diameter of the experimental dots in Fig. 2 of Ref. \cite{Blake2007}. From Fig. 2 and Table I the Fresnel-based model is clearly superior to the slab-based one. It seems anyway that the predictions of the spectral positions of the maxima for h = 300 nm and h = 90 nm in Fig. 2 are better guessed by the slab-based model. This is also the reason why $\tilde{\chi}_{0F}^2$ is not good for h = 300 nm. From simulations it turns out that the positions of the maxima are more sensitive to the substrate parameters than to the graphene layer; this is a possible explanation of the discrepancy from the Fresnel-based fit.       

The last graph of Fig. 2 considers a free standing graphene layer and compares $R_{s(p)}+T_{s(p)}$ for the present theory and for the slab model used in \cite{Blake2007}. The two theoretical predictions are very different. For a constant value of $\sigma=e^{2}/4\hbar$ the Fresnel theory predicts a constant absorption as a function of the wavelength whereas the slab model predicts a wavelength dependance. The two theoretical predictions are compared with the experimental data published in \cite{Nair2008}. In this case it was possible to retrieve $s$ (Table I) from the spreading of the experimental data. Again the superiority of the Fresnel model is quite evident. 
\begin{figure}
\includegraphics{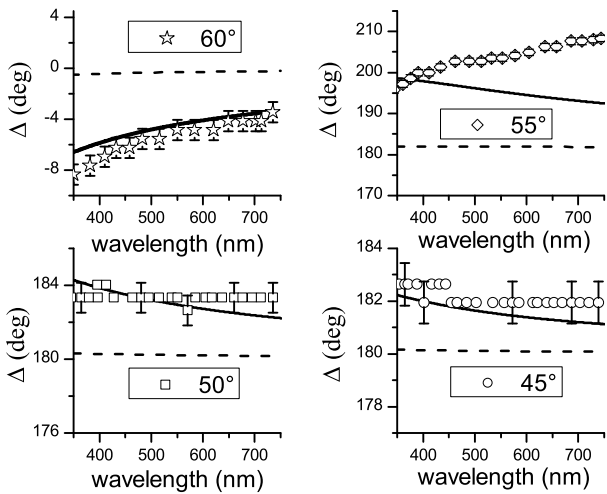}
\caption{\label{} Spectral dependance of the ellipsometric parameter $\Delta$ for graphene on an amorphous quartz substrate for each of the four $\theta_{i}$ considered in \cite{Kravets2010}. Experimental data are represented by dots, slab model predictions by a dashed line and Fresnel model predictions by a solid line.}
\end{figure}

To fix $\chi$, I turned my attention to spectroscopic ellipsometry of graphene flakes located on a flat amorphous quartz. Figures 3, 4, and 5 compare the theoretical predictions for the Fresnel-based model and for the slab-based model with the measurements published in \cite{Kravets2010}. Figures 3 and 4 show the simulated spectral dependence of the ellipsometric parameter $\Delta$ \cite{Tompkins} at four angles of incidence. Figure 3 shows the ensemble of the experimental data and the fits of the two models for $\theta_{i}$ =60$^\circ$. For clarity reasons Fig. 4 focuses on each $\theta_{i}$. From Figs. 3 and 4, for $\lambda>$ 350 nm, a good Fresnel fit is obtained for $\sigma=e^{2}/4\hbar$ and $\chi=1.0\cdot10^{-9}$ m. This is compared with the slab-based model used in \cite{Kravets2010} in the same frequency range. For $\theta_{i}$ =55$^\circ$ $\Delta$ does not fit very well, maybe because it is too close to Brewster and cross polarization effects \cite{Aiello09} may be present ($10^{-3}<R_{p}<1.5\cdot10^{-3}$ at $\theta_{i}$ =55$^\circ$ in this spectral range). From Fig. 4 and Table 1 the Fresnel-based model fits much better than the slab-based model the experimental data for the ellipsometric parameter $\Delta$, and hence it better predicts the phase of the reflected light. As for the measurements of contrast the standard deviations $s$ for the ellipsometric measurements have been extracted from the experimental line widths in Fig. 5(b) of Ref. \cite{Kravets2010}.  

The ellipsometric parameter $\Delta$ is very sensitive to the graphene film because $\Delta$ = 180 $^\circ$ or 0$^\circ$ for the quartz substrate and all of the non trivial phase contribution to the reflection coefficients comes from graphene. In particular from formulas (6) the phase of the reflected light is different from $0^\circ$ or $180^\circ$ only if $\chi\neq0$ m. The experimental data for $\Delta$ allow one to fix a value of $\chi$ within $\pm 1 \cdot 10^{-10}$ m, and they are also sensitive to its sign. 

Figure 5 shows the ellipsometric parameter $\Psi$ \cite{Tompkins}. From Fig. 5 and Table I the Fresnel fit and the slab fit for $\Psi$ give almost equivalent predictions, with a slight superiority of the slab-based one. In this case the role of the dielectric substrate is overwhelming (more than $98\%$ of the $\Psi$ signal is due to it). This is a possible explanation of the small discrepancy of the experimental data from the Fresnel-based fit. The fit reported in Fig.5 is for the same values used to fit $\Delta$. A better fit is obtained if we let $\sigma$ be 30$\%$ or $40\%$ less than this value. This does not really affect $\Delta$ but it can be an indication of defects in the sample. In fig 5 dots have the linear dimensions of one standard deviation $s$ (experimental line width in Fig. 5(a) of Ref. \cite{Kravets2010}), this is why error bars are not reported.             
\begin{figure}
\includegraphics{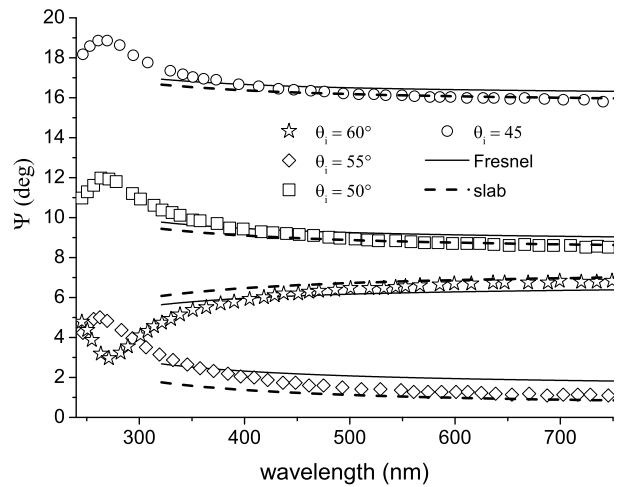}
\caption{\label{} Spectral dependance of ellipsometric parameter $\Psi$ for graphene on an amorphous quartz substrate. Experimental data for the four $\theta_{i}$ are extracted from \cite{Kravets2010}. The Fresnel and the slab model simulation for each $\theta_{i}$ and for $\lambda$ $>$ 350 nm are reported. Fits are obviously relative to their nearest experimental data points.}
\end{figure}

For $\lambda$ $<$ 350 nm data indicate that $\sigma$ and $\chi$ vary sensibly with $\lambda$. In this spectral range it is possible for each wavelength to fix a different value of $\sigma$ and $\chi$. We consider for instance $\lambda$ = 270 nm. At this wavelength by setting $\sigma=3.1\cdot e^{2}/4\hbar$ and $\chi=-1.2\cdot10^{-9}$ m we obtain for $\theta_{i}$ = 60$^\circ$, 55$^\circ$, 50$^\circ$, 45$^\circ$ respectively $\Psi$ = 3.2$^\circ$, 4.8$^\circ$, 12.1$^\circ$, 18.9$^\circ$ and $\Delta$ = 18.2$^\circ$, 168$^\circ$, 175$^\circ$, 177.5$^\circ$,  and an absorption of 6.5\%, in very nice agreement with data in Ref. \cite{Kravets2010}. From simulations: $|\chi|$ is still too small to influence absorption and so it is fixed by $\sigma$, and $\chi$ fixes the rest; $\Delta$ is sensitive to the sign of $\chi$, $\Psi$ to its magnitude only. A negative value of $\chi$ at 270 nm is probably related to the fact that the real part of the dielectric constant of graphite become negative around 4 eV and up to around 7 eV \cite{Taft64}. Fresnel coefficients simulate as well spectroscopic ellipsometry of graphene deposited on Si$\rm{O_{2}}$/Si reported in the same paper \cite{Kravets2010}. I confirm that $\Psi$ is sensitive to presence of the graphene layer. Unfortunately (even up to $|\chi|=5\cdot10^{-9}$ m) $\Delta$ is dominated by the signal from the substrate.

In conclusion the comparison with the experimental results shows that the Fresnel coefficients are essential to interpret the most remarkable experiments in graphene optics. Any hypothesis on an effective thickness of a single-layer 2D atomic crystal as required by modeling it as an homogeneous slab is not necessary. This last model is not able to reproduce properly either the absorption of graphene, or the phase of its reflection coefficient. On this basis any physical parameter, deduced from it, is hardly meaningful. As for bulk materials, ellipsometry is able to furnish both $\chi$ and $\sigma$, showing that these are the physically meaningful quantities experimentally accessible from the linear optical response of a 2D atomic crystal. In particular for graphene, from the ensemble of the experimental data considered, in the spectral range 450 nm $<\lambda<$ 750 nm, $\sigma=6.08\cdot 10^{-5}\pm 2\cdot10^{-5} \Omega^{-1}$ and $\chi=8\cdot 10^{-10}\pm 3\cdot 10^{-10}$ m. Of course the Fresnel coefficients are not valid for 2D crystal bilayers or multilayers. In these cases light propagates from one layer of atoms to the other and a thickness should be considered. This shows once more how special single-layer atomically thin 2D crystals are.

\begin{table}
\caption{Chi-squared test value for the Fresnel-based model fit and the slab-based model fit; $n$: number of experimental data points considered for the fitting procedure.}
\begin{tabular}{c c c c c c c}
\hline
 & $n$ & $s$ & $\tilde{\chi}_{0F}^2$ & $P_{F}$ & $\tilde{\chi}_{0S}^2$ &  $P_{S}$ \\ 
\hline
Contrast 300 nm & 12 & 0.01 & 4.35 & $< 0.1\% $ & 4.93 & $ < 0.1\%$ \\ 
Contrast 200 nm & 12 & 0.01 & 0.15 & $99.8\% $ & 1.68 & $  9.7\%$ \\
Contrast 90 nm & 12 & 0.01 & 1.18 & $30.2\% $ & 6.42 & $ < 0.1\%$ \\
Absorption & 34 & 0.0015 & 1.09 & $33.4 \%$ & 6.75 & $ < 0.1\%$ \\
$\Delta$ \quad $60^\circ$ & 25 & $0.8^\circ$ & 1.60 & $5\% $ & 45.74 & $ < 0.1\%$ \\ 
$\Delta$ \quad $55^\circ$ & 20 & $0.8^\circ$ & 59.87 & $< 0.1\% $ & 333.37 & $ < 0.1\%$ \\ 
$\Delta$ \quad $50^\circ$ & 23 & $0.8^\circ$ & 0.82 & $69\% $ & 18.42 & $ < 0.1\%$ \\ 
$\Delta$ \quad $45^\circ$ & 28 & $0.8^\circ$ & 0.55 & $94\% $ & 7.74 & $ < 0.1\%$ \\ 
$\Psi$ \quad $60^\circ$ & 29 & $0.4^\circ$ & 0.98 & $49\% $ & 1.92 & $  0.4\%$ \\ 
$\Psi$ \quad $55^\circ$ & 29 & $0.4^\circ$ & 2.21 & $< 0.1\% $ & 2.48 & $  < 0.1\%$ \\
$\Psi$ \quad $50^\circ$ & 29 & $0.4^\circ$ & 0.96 & $53\% $ & 0.77 & $ 79\%$ \\
$\Psi$ \quad $45^\circ$ & 29 & $0.4^\circ$ & 0.78 & $77\% $ & 0.47 & $ 98\%$ \\

\hline 
\end{tabular}
\end{table}

\section{ACKNOWLEDGMENTS}
I acknowledge Luca Dell'Anna and Nicola Lo Gullo for useful discussions.

\bibliography{letter}
\end{document}